# Giant Electroresistance in Ferroelectric Tunnel Junctions

M. Ye. Zhuravlev[1], R. F. Sabirianov,[2,3] S. S. Jaswal,[1,3] and E. Y. Tsymbal[1,3*]

[1]*Department of Physics and Astronomy, University of Nebraska, Lincoln, NE 68588-0111*
[2]*Department of Physics, University of Nebraska, Omaha, NE 68182-0266*
[3]*Center for Materials Research and Analysis, University of Nebraska, Lincoln, NE 68588-0111*

The interplay between the electron transport in metal/ferroelectric/metal junctions with ultrathin ferroelectric barriers and the polarization state of a barrier is investigated. Using a model which takes into account screening of polarization charges in metallic electrodes and direct quantum tunneling across a ferroelectric barrier we calculate the change in the tunneling conductance associated with the polarization switching. We find the conductance change of a few orders of magnitude for metallic electrodes with significantly different screening lengths. This giant electroresistance effect is the consequence of a different potential profile seen by transport electrons for the two opposite polarization orientations.

PACS: 73.40.-c, 73.40.Gk, 77.80.Fm, 85.50.Gk

In recent years, ferroelectric materials have attracted significant interest because of their promising potential in various technological applications.[1,2] For example, due to their spontaneous dielectric polarization that can be switched by an applied electric field, ferroelectrics can be used as binary data storage media in nonvolatile random access memories. Recent experimental and theoretical findings suggest that ferroelectricity persists down to vanishingly small sizes, which opens a possibility to further miniaturize electronic devices based on ferroelectric materials.[3] In particular, it was discovered that, in organic ferroelectrics, ferroelectricity can be sustained in thin films of a few monolayer thickness.[4] In perovskite ferroelectric oxides, ferroelectricity was observed down to a nanometer scale.[5] This fact is consistent with first-principle calculations that predict a nanometer critical thickness for a perovskite ferroelectric film sandwiched between two metals.[6] The existence of ferroelectricity at such a small film thickness makes it possible to use ferroelectrics as *tunnel* barriers in metal/ferroelectric/metal (M/FE/M) junctions. Recent experiments indicate that the electrical resistance in M/FE/M junctions with ultrathin barriers depends on the orientation of the dielectric polarization which can be switched by an applied electric field.[7] The origin of this electroresistance effect is not completely understood and to the best of our knowledge no modeling of this phenomenon has been performed.

In this Letter, using a simple model for an ultrathin ferroelectric (FE) barrier separating two different metal electrodes ($M_1$ and $M_2$), we investigate the electroresistance effect in ferroelectric ($M_1$/FE/$M_2$) tunnel junctions. We show that the reversal of the dielectric polarization in the ferroelectric produces a change in the electrostatic potential profile across the junction. This leads to the resistance change which can reach a few orders of magnitude for metal electrodes with significantly different screening lengths. We designate this phenomenon as the *giant electroresistance* (GER) effect.

The physical mechanism which is responsible for the GER in ferroelectric tunnel junctions (FTJs) is the change of the electrostatic potential profile $\varphi(z)$ induced by the reversal of the dielectric polarization **P** in the ferroelectric. Indeed, if the ferroelectric film is sufficiently thin but still maintains its ferroelectric properties the surface charges in the ferroelectric are not completely screened by the adjacent metals and therefore the depolarizing electric field **E** in the ferroelectric is not zero.[2] The electrostatic potential associated with this field depends on the direction of the dielectric polarization. If a FTJ is made of metal electrodes which have different screening lengths, this leads to the asymmetry in the potential profile for the opposite polarization directions. Thus, the potential seen by transport electrons changes with the polarization reversal which leads to the GER effect.

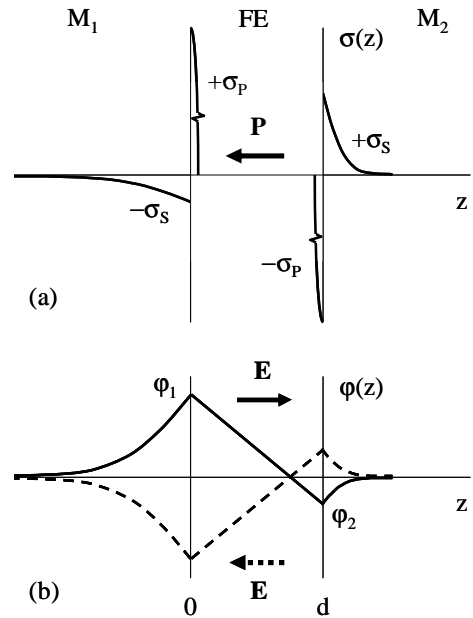

**Fig.1** Electrostatics of a $M_1$/FE/$M_2$ junction: (a) charge distribution and (b) the respective electrostatic potential profile (solid line). It is assumed that metal 1 ($M_1$) and metal 2 ($M_2$) electrodes have different screening lengths ($\delta_1 > \delta_2$) which lead to the asymmetry in the potential profile. The dashed line in (b) shows the potential when the polarization **P** in the ferroelectric is switched resulting in the reversal of the depolarizing field **E**.

In order to make these arguments quantitative we consider a ferroelectric thin film of thickness $d$ placed between two different semiinfinite metal electrodes. The ferroelectric is polarized in the direction perpendicular to the plane. The polarization **P** creates surface charge densities, $\pm\sigma_P = \pm|\mathbf{P}|$, on the two surfaces of the ferroelectric film. These polarization charges, $\pm\sigma_P$, are screened by the screening charge per unit area, $\mp\sigma_S$, which is induced in the two metal electrodes, as is shown schematically in Fig.1a. We assume that the ferroelectric is perfectly insulating so that all the compensating (screening) charge resides in the electrodes. Further, we assume that the FTJ junction is short-circuited, that is connected to a low-impedance source, which equalizes the potentials of the two electrodes at infinity. In order to find the distribution of the screening charge and the potential profile across the junction, we apply a Thomas-Fermi model of screening (e.g., ref. [8]). According to this model the screening potential within metal 1 ($z \leq 0$) and metal 2 ($z \geq d$) electrodes is given by

$$\varphi(z) = \begin{cases} \dfrac{\sigma_S \delta_1 e^{-|z|/\delta_1}}{\varepsilon_0}, & z \leq 0 \\ -\dfrac{\sigma_S \delta_2 e^{-|z-d|/\delta_2}}{\varepsilon_0}, & z \geq d \end{cases} \quad (1)$$

Here $\delta_1$ and $\delta_2$ are the Thomas-Fermi screening lengths in the $M_1$ and $M_2$ electrodes and $\sigma_S$ is the magnitude of the screening charge per unit area which is to be the same in metals 1 and 2 due to the charge conservation condition. Note that the short circuit condition has been included in Eqs.(1), which follows from the fact that $\varphi(z) \to 0$ when $z \to \pm\infty$. The screening charge $\sigma_S$ can be found from the continuity of the electrostatic potential, implying that the potential drop within the ferroelectric film is determined by a constant electric field in the ferroelectric:

$$\varphi(0) - \varphi(d) = \frac{d(P - \sigma_S)}{\varepsilon_F}. \quad (2)$$

We note that here $P$ is considered to be the absolute value of the spontaneous polarization, and the introduction of the dielectric permittivity $\varepsilon_F$ is required to account for the induced component of polarization resulting from the presence of an electric field in the ferroelectric. Now using Eqs. (1), (2) and introducing the dielectric constant $\varepsilon = \varepsilon_F/\varepsilon_0$ we arrive at

$$\sigma_S = \frac{dP}{\varepsilon(\delta_1 + \delta_2) + d}. \quad (3)$$

It is evident from Eq.(3) that for "good" metals in which the screening length is small (a fraction of an Angstrom) and for not too thin ferroelectrics, such that $\dfrac{\varepsilon(\delta_1 + \delta_2)}{d} \ll 1$, a full screening occurs, i.e. $\sigma_S = P$, which implies no depolarizing field **E** in the ferroelectric. In the opposite limit, $\dfrac{\varepsilon(\delta_1 + \delta_2)}{d} \gg 1$, the screening charge tends to zero and the depolarizing field increases to saturation at $\mathbf{E} = -\mathbf{P}/\varepsilon$.[2]

Fig. 1b shows the electrostatic potential in a $M_1$/FE/$M_2$ junction assuming that metals $M_1$ and $M_2$ have different screening lengths, such that $\delta_1 > \delta_2$. It follows from Eq. (1) that different screening lengths result in different absolute values of the electrostatic potential at the interfaces, so that $\varphi_1 \equiv |\varphi(0)| \neq \varphi_2 \equiv |\varphi(d)|$, which makes the potential profile highly asymmetric, as is seen from Fig.1b.[9] The switching of the polarization in the ferroelectric layer leads to the change in the potential which transforms to the one shown in Fig.1b by the dashed line. Thus, due to different screening lengths in the two metals that make the electrostatic potential profile asymmetric, the switching of the polarization orientation in the ferroelectric barrier should inevitably lead to the change in the resistance of the junction.

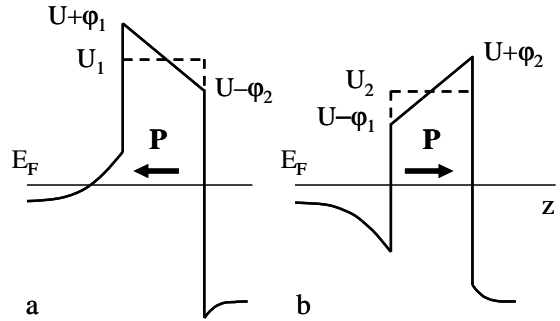

**Fig.2** Potential profile $V(z)$ in a $M_1$/FE/$M_2$ junction for polarization pointing to the left (a) and for polarization pointing to the right (b) assuming that $\delta_1 > \delta_2$. Values of the potential at the interfaces with respect to the Fermi energy are displayed. The dashed lines show the average potential seen by transport electrons tunneling across the ferroelectric barrier. The horizontal solid line denotes the Fermi energy, $E_F$.

In order to predict the magnitude of the resistance change associated with polarization switching we assume that the thickness of the ferroelectric barrier is so small that the dominant transport mechanism across the FTJ is the direct quantum-mechanical electron tunneling. The overall potential profile $V(z)$ seen by tunneling electrons is a superposition of the electrostatic potential shown in Fig.1b and the potential barrier created by the ferroelectric insulator. For simplicity we assume that the latter has a rectangular shape of height $U$ with respect to the Fermi energy.[10] The electronic potentials within the metal electrodes are determined by the screening lengths $\delta_1$ and $\delta_2$ which are related to the Fermi wave vectors, $k_{1,2}$, according to the Thomas-Fermi theory, by $k_{1,2} = \frac{\pi a_0}{4\delta_{1,2}^2}$, where $a_0$ is the Bohr radius.[8] The resulting potential $V(z)$ for the two opposite orientations of polarization in the ferroelectric barrier is shown schematically in Fig.2 for $\delta_1 > \delta_2$.

At a small applied bias voltage the conductance of a tunnel junction per area $A$ is obtained using the standard expression:[11]

$$\frac{G}{A} = \frac{2e^2}{h} \int \frac{d^2 k_\parallel}{(2\pi)^2} T(E_F, \mathbf{k}_\parallel), \quad (4)$$

where $T(E_F, \mathbf{k}_\parallel)$ is the transmission coefficient evaluated at the Fermi energy $E_F$ for a given value of the transverse wave vector $\mathbf{k}_\parallel$. The transmission coefficient is obtained from the Schrödinger equation for an electron moving in the potential $V(z)$ by imposing a boundary condition of the incoming plane wave normalized to unit flux density and by calculating the amplitude of the transmitted plane wave. We assume for simplicity that electrons have a free electron mass in all the three layers. The Fermi energy in metal 2 is fixed at $E_F = 3.5$ eV (with respect to the bottom of the band), resulting in the screening length of $\delta_2 \approx 0.07$ nm typical for a "good" metal. The potential barrier is assumed to be $U = 0.5$ eV typical for a ferroelectric insulator.[7] The dielectric constant of the ferroelectric is assumed to be $\varepsilon = 2000$ which is a representative value for perovskite ferroelectrics.[1]

Fig.3a shows the calculated amplitudes of the potential $\varphi_1 \equiv |\varphi(0)|$ and $\varphi_2 \equiv |\varphi(d)|$ at the M$_1$/FE and M$_2$/FE interfaces as a function of the screening length, $\delta_1$, in the M$_1$ electrode. The difference between $\varphi_1$ and $\varphi_2$ controls the asymmetry in the potential profile which is decisive for the resistance change on polarization switching. Indeed, the average potential barrier height seen by transport electrons traveling across the ferroelectric layer for polarization pointing to the left, $U_1 = U + \varphi_1 - \varphi_2$, is not equal to the average potential barrier height for polarization pointing to the right, $U_2 = U - \varphi_1 + \varphi_2$, as is seen from Figs.2a,b. It follows from Fig.3a that a relatively large screening length in the M$_1$ layer ($\delta_1 \gg \delta_2$) leads to $\varphi_1 \gg \varphi_2$ and, hence, to $U_1 > U_2$. This makes the conductance $G_1$ for polarization pointing to the left much smaller than the conductance $G_2$ for polarization pointing to the right, thereby resulting in the GER effect.

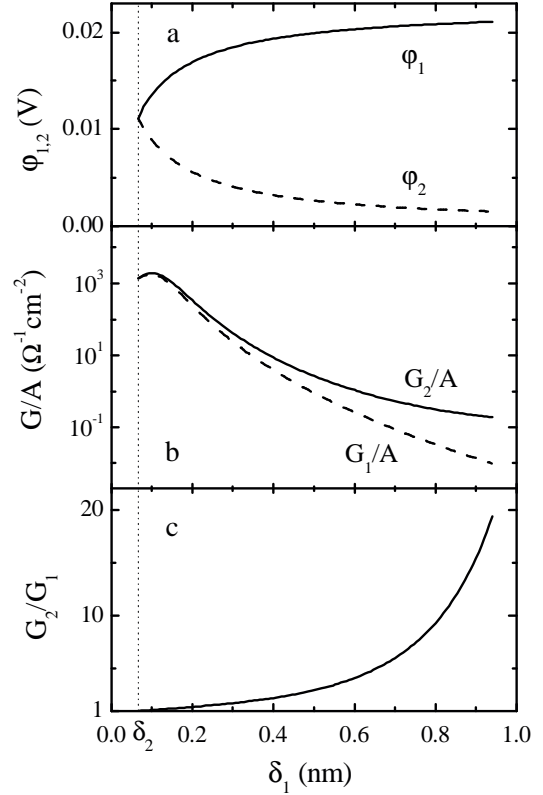

**Fig.3** Calculated data as a function of screening length, $\delta_1$, in metal 1 electrode for $P=20$ μC/cm$^2$ and $d=2$ nm: (a) amplitudes of the potential at the M$_1$/FE (solid line) and M$_2$/FE (dashed line) interfaces; (b) conductance per unit area for polarization oriented to the right, $G_2/A$ (solid line) and for polarization oriented to the left, $G_1/A$ (dashed line); conductance change, $G_2/G_1$, associated with the polarization switching in the ferroelectric barrier. The vertical dotted line indicates the value of $\delta_1 = \delta_2$ at which no asymmetry in the potential profile and, hence, no conductance difference is predicted.

The latter fact is evident from the calculated conductance values per unit area, $G_1/A$ and $G_2/A$, shown in Fig.3b. For $\delta_1 = \delta_2$ there is no asymmetry in the potential ($\varphi_1 = \varphi_2$) and therefore $G_1 = G_2$. With increasing $\delta_1$ both $G_1/A$ and $G_2/A$ decrease reflecting the drop in the Fermi wave vector $k_1$. This decrease is accompanied by the departure of the $G_1/A$ and $G_2/A$ curves from each other. The figure of merit is the degree of the conductance (resistance) change in response to the polarization reversal, which we define by the GER ratio, $G_2/G_1$, shown in Fig.3c. It is seen that with increasing the screening length this ratio increases exceeding factor of 10 when $\delta_1$ approaching 1 nm. Our calculation predicts a further increase in the GER with $\delta_1$ even when the potentials $\varphi_1$ and $\varphi_2$ are close to saturation. This is the consequence of the increasing effective thickness of the tunneling barrier for the case when polarization points to the left, as is seen in Fig.2a. The latter is due to the electrostatic potential $\varphi_1$ at the $M_1$/FE interface exceeding the Fermi energy in metal 1 electrode which occurs, for the parameters chosen, when $\delta_1$ is greater than 0.25nm. For $\delta_1$=0.6nm, which is the approximate screening length calculated from first-principles for $SrRuO_3$ metal,[6] the GER ratio is $G_2/G_1 \approx 4$. This result is consistent with the resistance change obtained for $SrRuO_3/Pb(Zr_{0.52}Ti_{0.48})O_3/Pt$ junctions,[7] though these junctions might not be in the true direct tunneling regime.

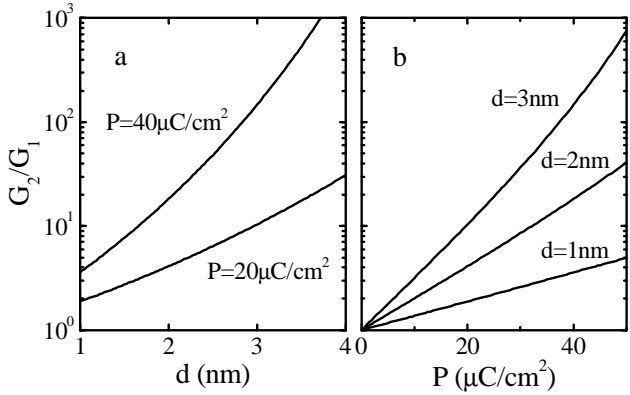

**Fig.4** GER ratio, $G_2/G_1$, as a function of ferroelectric barrier thickness, $d$, for different values of polarization (a) and as a function of polarization, $P$, for different barrier thicknesses (b). $\delta_1$=6Å.

Fig.4a shows the GER ratio as a function of ferroelectric layer thickness. The increase in $G_2/G_1$ with $d$ evident from this figure is the consequence of a different effective potential barrier height for the two polarization orientations. Indeed, as follows from Fig.2a, the average potential seen by tunneling electrons for the polarization pointing to the left, $U_1$, is higher than that for the reversed polarization, $U_2$. Therefore, the conductance $G_1$ decays faster than $G_2$ with increasing thickness $d$, causing the GER to increase exponentially with the thickness. As is seen from Fig.4a, the higher polarization value enhances both the GER and the degree of its change with the thickness resulting from a higher potential difference for opposite polarizations.

This conclusion is consistent with the dependence of the GER ratio versus polarization which is shown in Fig.4b. As is seen from this figure, $G_2/G_1$ increases with $P$ exponentially being enhanced for thicker barriers.

In conclusion, we have demonstrated the possibility and explained the mechanism of giant electroresistance in ferroelectric tunnel junctions. Using a model which takes into account screening of polarization charges and direct quantum tunneling across the ferroelectric barrier we calculated the change in the tunneling conductance associated with the switching of polarization in the ferroelectric. For metal electrodes with significantly different screening lengths, we found that the conductance can change by a few orders of magnitude reflecting the different potential profile seen by transport electrons for the two opposite polarization orientations. These results are encouraging in view of potential applications of ferroelectric tunnel junctions as binary data storage media in nonvolatile random access memories. We hope that our theoretical predictions will stimulate experimental studies of the giant electroresistance effect in ferroelectric tunnel junctions.

This work is supported by NSF (grants DMR-0203359 and MRSEC: DMR-0213808) and the Nebraska Research Initiative.

* Electronic address: tsymbal@unl.edu